\documentclass[reprint,aip,graphicx]{revtex4}
\usepackage{graphicx}% Include Figure files
\usepackage{dcolumn}% Align table columns on decimal point
\usepackage{bm}% bold math
\usepackage{amssymb}
\usepackage{color}
\usepackage{caption}
\usepackage{epstopdf}
\begin{document}
\title{Mode Localization in the Cooperative Dynamics of Protein Recognition}
\author{J. Copperman}
\affiliation{Department of Physics, University of Oregon, Eugene, Oregon 97403}
\author{M.G. Guenza\footnote{Author to whom correspondence should be addressed. Electronic mail: mguenza@uoregon.edu}}
\affiliation{Department of Chemistry and Institute of Theoretical Science, University of Oregon, Eugene, Oregon 97403}
\date{\today}

\begin{abstract}
The biological function of proteins is encoded in their structure and expressed through the mediation of their dynamics. Local fluctuations are known to initiate biologically relevant pathways as they cooperatively enhance the dynamics in specific regions in the protein. Those biologically active regions provide energetically-comparable conformational states that can be trapped by a reacting partner. We analyze this mechanism as we calculate the dynamics of monomeric and dimerized HIV protease, and free Insulin Growth Factor II Receptor (IGF2R) domain 11 and its IGF2R:IGF2 complex. We adopt a newly developed coarse-grained model, the Langevin Equation for Protein Dynamics (LE4PD), which predicts dynamical relevant mechanisms with high accuracy. Both simulation-derived and experimental NMR conformers are the input structural ensembles for the LE4PD. The use of the experimental NMR conformers requires minimal computational resources.
\end{abstract}

\maketitle
\section{Introduction}
The biological function of proteins is uniquely defined by the protein primary sequence, which determines its three-dimensional structure and dynamics.\cite{pettitt1990advances} The protein local motion develops along a pathway of transitions between metastable states inside a hierarchy of structures that, on the local scale, are separated by small energy barriers of the order of $k_BT$. It is the unique structure of this configurational landscape that determines the specific thermodynamic and kinetic properties of a protein and defines its ability of regulating its function. 

Residue-specific, localized dynamics are relevant in the energetic pathways of binding.\cite{matoba2003atomic,dodson2008molecular} Pre-existing pathways in the free-energy landscape have been found to guide the transmission of the allosteric signals.\cite{boehr2009role} It has also been shown that point mutations of residues, and also small ligand binding, can lead to an identical folded configuration while they modify the protein biological activity. This indicates that, in those cases, not the structure but the dynamics has the dominant effect on the protein function.\cite{popovych2006dynamically} Protein dynamics can facilitate docking of a ligand.\cite{mobley2009binding} Often the bound conformation of the protein is different from its apo structure: in those cases the bound conformation populates a preexisting configuration that is just less probable than the apo structure, but should be visible in a dynamic study.\cite{okazaki2008dynamic} 

It is well established that local fluctuations largely affect protein reactivity; however a simple theoretical model that directly and precisely relates structure to dynamics is still needed.\cite{tzeng2012protein} The Langevin Equation for Protein Dynamics (LE4PD) is a quantitative model of protein dynamics and directly relates the protein structure to its local fluctuations. Because the LE4PD precisely models local dynamics, it is used here to study the role of fluctuations in the biological activity of a number of sample proteins. 

Starting from the static structural ensemble of a protein, the LE4PD predicts where specific regions in the protein three-dimensional structure become dynamically active at a given timescale and how allosteric dynamics are enhanced or suppressed upon binding. The extent of the energetic barriers involved indicates if a new dynamically active region emerges upon binding that is likely involved in the following step in a reaction pathway, or if entropically-relevant multiple states emerge that are distributed along the protein for a straightforward entropy-enthalpy compensation mechanism. As an example we study both monomer and dimerized HIV protease,\cite{ishima2001folded,ishima2003solution} and free Insulin Growth Factor II Receptor (IGF2R) domain 11 and the IGF2R:IGF2 complex.\cite{williams2007structural}

Other theoretical models based upon effective harmonic descriptions, which calculate sequence dependent protein flexibility, are Normal Mode Analysis (NMA), and general Elastic Network Models (ENM) including the Gaussian Network Model (GNM).\cite{go1983dynamics,bahar1997direct,atilgan2001anisotropy,yang2007well,yang2009protein} Differently than the LE4PD, those models were designed to study short-time vibrational fluctuations around the protein structure, which are dominated by the topology of native contacts. Because the LE4PD is a diffusive equation of motion which contains information about the extent of the intramolecular energy barriers, specific monomer friction coefficient, amino-acid specific local semiflexibility, degree of hydrophobicity, as well as hydrodynamics, it provides a realistic description of the motion of proteins in solution over a wide range of timescales, from the local vibrational fluctuations as measured by crystallographic B-factors, to the long-time dissipative dynamics. In the short time regime elastic network models and the LE4PD predict qualitatively consistent dynamics.\cite{caballero2007theory}

For a folded protein a number of configurational states are available in spatially well-defined regions along the primary sequence. At a given temperature a part of the protein, for instance a loop or a tail, may be intrinsically disordered and populating a number of thermally activated conformational states that are metastable, energetically similar, and so equally probable. Enhanced fluctuations in the spatial positions of key residues or short fragments can make possible the trapping of favorable configurations by a reactant or a substrate, following the well-established conformational selection model of binding.\cite{monod1965nature} When the correct local configuration for binding is available in the configurational landscape of the isolated protein, the extent of free energy needed for binding is reduced. As the trapping of a favorable state does not require overcoming an energy barrier, other processes, for example inter-diffusion of the reaction partners, become the slow relevant processes that determine the rate of the reaction. This conformational selection process for the recognition event, can be followed by a relaxation or induced fit to the bound conformation.\cite{grunberg2006flexibility} 

The gain in energy as the protein transitions over the energy barrier and reaches the bound state has to be small to allow the possible breaking of the reaction product when new conditions arise that are destabilizing this state: the process needs to be flexible enough to permit the progress towards the following reaction step without dramatic gains or loss of free energy. In this delicate balance of energy, modulated within an energy window of a few $k_B T$, the primary sequence of a protein plays a decisive role. Emergentsequence-dependent fluctuations of isolated and bound proteins are predicted by our approach, which is a coarse-grained but still physically realistic representation of the motion of biological macromolecules at the lengthscale of a single amino acid and larger. 

Our model provides information on the nature of the conformational energy barriers involved in the binding process, and on the length scale and time scale of the dynamical fluctuations involved in binding. The presence of barriers and their height play an important role in the binding reaction. Transitions need to be energetically activated to render the biological process forbidden if the temperature lowers below physiological conditions. However, the barriers need to be small to make their crossing possible at physiological temperature, as the dynamics are ``fueled'' by the thermal fluctuations of the surrounding liquid.\cite{koshland1958application,lewandowski2015direct} 

Experimentally it has been observed that upon binding the loss in entropy of the protein, which would oppose the reaction, is often paired to the emergence of disorder in remote regions of the protein, apparently uncorrelated to the binding site.\cite{grunberg2006flexibility} New flexible regions often arise in the relaxed bound state, or exposed hydrophobic residues are found to transition to the hydrophobic region, becoming protected and increasing the entropy of the solvent in the well-known mechanism of enthalpy-entropy compensation.\cite{lumry1970enthalpy,dunitz1995win}

In its diffusive mode description, the LE4PD allows for the identification of a variety of dynamical processes that emerge at increasing timescales.  The diffusive mode solution of the LE4PD organizes the configurational landscape in a linearly independent set of variables. Fluctuations are defined  on a range of length and timescales, with dynamics encompassing the relative motion between neighboring $\alpha$-carbons to the global rotations of the structure as a whole.\cite{caballero2007theory,copperman2014coarse-grained} In this study internal modes of motion that present energetically activated local dynamics are identified through this model, together with the characteristic length and timescales of their dynamics. A range of equilibrium dynamical processes emerge on different timescales following a hierarchical scheme, suggesting a possible sequential mechanism in the non-equilibrium reaction pathway.

The diffusive mode rendition precisely indicates the position inside the primary sequence of these energetically-guided local fluctuations, and provides information about the extent of localization of these activated dynamics. This indicates if the motion involves a single residue or a number of cooperatively moving specific residues. By identifying and analyzing the regions of local flexibility and cooperative motion of the residues inside a protein, we argue that it is possible to learn which parts of the protein will lead the kinetics of the biologically relevant processes. In this way, the LE4PD model provides a straigthforward and visually intuitive representation of the locations of enhanced reactivity, or ``binding regions,'' and the emergent length and timescales of motion. For all the proteins in this study, the LE4PD method clearly indicates regions of high mobility and slow, large-amplitude dynamics, which are trapped and not detected in the bound forms of the protein. These regions directly correspond to regions with highly conserved residues in the family of proteins with related biological function, and are directly involved in the binding interactions. 

Input to the LE4PD is an ensemble of structural configurations, which has to be representative of the distribution of folded states of the protein. While proteins sample a very large  $3 N$-dimensional configurational space, with $N$ the number of independent sites comprising the protein, at the bottom of the funnel-like energy landscape the conformational diversity is much smaller.\cite{bryngelson1995funnels,onuchic1997theory,dill1997levinthal} A common paradigm is that the important internal fluctuations of a folded protein span a limited number of specific structures,\cite{monod1965nature,levinthal1968there} and these can be well sampled  experimentally by NMR.\cite{lange2008recognition} Overall we observe that ensembles generated by the experimental NMR solution structure ensembles, and that generated from atomistic molecular dynamics simulation, lead to consistent and accurate results. The LE4PD with input from NMR ensembles requires very limited computational resources.

\section{Theoretical Approach: the Langevin Equation for Protein Dynamics}
\label{theory}
The LE4PD is a microscopic  theory based on the Rouse-Zimm dynamics of synthetic macromolecules in solution and modified to properly describe protein dynamics.\cite{bixon1978optimized,doi1986theory} While synthetic polymers are mostly isotropic in shape and unstructured as well as fully exposed to the solvent, proteins are anisotropic in shape and have a hydrophobic core with amino acids that are only partially exposed to solvent, depending on their position in the three-dimensional structure.\cite{caballero2007theory,copperman2014coarse-grained} 

The protein is described as a sequence of coarse-grained units, each centered on the $\alpha$-carbon of each amino acid. Their dynamics are driven by the balance of viscous dissipation and the entropic restoring force. The effect of the solvent enters the dynamics as friction and Brownian forces due to the random collisions of the fast-moving particles, whose dynamics are projected onto the coordinates of the coarse-grained units. In this formalism the solvent is described as an effective medium.

The orientational  Langevin equation governing the bond dynamics is 
\begin{eqnarray}
\frac{\partial\vec{l}_i (t)}{\partial t}= - \sigma  \sum_{j,k} L_{ij}U_{jk}\vec{l}_k(t)+\vec{v}_i(t) \ ,
\label{LE}
\end{eqnarray}
 with $i,j=1, ..., N-1$. The overall timescale is set by $\sigma=3 k_B T/(l^2\overline{\zeta})$ where $k_B$ is the Boltzmann constant, $T$ is the temperature, $l^2$ is the squared bond distance, and $\overline{\zeta}$ is the average monomer friction coefficient. $\vec{v}_i(t)$ is the random delta-correlated bond velocity. The matrix $\mathbf{L}$ is related to the preaveraged hydrodynamic interaction matrix, which describes the interaction between protein sites propagating through the liquid, described as a continuum medium, and partially screened by the dense hydrophobic protein core.\cite{caballero2007theory} The $\mathbf{U}$ matrix is the bond correlation matrix with $(\mathbf{U}^{-1})_{ij}=\frac{\langle \vec{l}_i \cdot \vec{l}_j \rangle}{\langle |\vec{l}_i | \rangle \langle | \vec{l}_j | \rangle }$.

%Eq. \ref{LE} represents a set of $N-1$ first-order coupled differential equations, which are solved by finding the matrix of eigenvectors $\mathbf{Q}$ which diagonalizes the product of matrices $\mathbf{LU}$. In these diffusive modes we have $N-1$ uncoupled linear diffusion equations where each mode is just a linear sum of the original bond vector basis $\vec{\xi_a}(t)=\sum_{i}Q^{-1}_{ai}\vec{l}_i(t)$. We define $\lambda_a$ to be the eigenvalues of $\sum_{i,j,k}Q^{-1}_{ai}L_{ij}U_{jk}Q_{kb}=\delta_{ab}\lambda_a$, ordered from smallest to largest $\lambda$, and $\mu_a$ as the eigenvalues of the bond correlation matrix alone $\sum_{i,j}Q^{-1}_{ai}U_{ij}Q_{jb}=\delta_{ab}\mu_a$. The $\mathbf{Q}$ matrix is a set of eigenvectors which spans the the $N-1$ dimensional space of bond vectors. Like the set of bond vectors $\vec{l}_i(t)$ the set of normal coordinates $\vec{\xi}_a(t)$ defines the instantaneous conformation of the macromolecule. The diffusive mode basis spans the same space as the bond vector basis with the added benefit of linearity: $\langle\vec{\xi}_a (t) \cdot\vec{\xi}_b (t) \rangle=\delta_{ab}  l^2/\mu_a$. A straightforward transformation relates the bond coordinates to the coordinates of the $\alpha$-carbon (see the Supplemental Information).

Eq. \ref{LE} represents a set of $N-1$ first-order coupled differential equations, which are solved by finding the matrix of eigenvectors $\mathbf{Q}$ which diagonalizes the product of matrices $\mathbf{LU}$. In these diffusive modes we have $N-1$ uncoupled diffusion equations where each mode is just a linear sum of the original bond vector basis $\vec{\xi_a}(t)=\sum_{i}Q^{-1}_{ai}\vec{l}_i(t)$. We define $\lambda_a$ to be the eigenvalues of $\sum_{i,j,k}Q^{-1}_{ai}L_{ij}U_{jk}Q_{kb}=\delta_{ab}\lambda_a$, ordered from smallest to largest $\lambda$, and the mode length  $\langle\vec{\xi}^2_a \rangle =  \frac{l^2}{\mu_a}$ with $\mu_a \equiv \sum_{i,j}Q^{-1}_{ai}U_{ij}Q_{ja}$. Like the set of bond vectors $\vec{l}_i(t)$ the set of modes $\vec{\xi}_a(t)$ defines the instantaneous configuration of the protein backbone, and a straightforward transformation relates the bond coordinates to the coordinates of the $\alpha$-carbon (see the Supplemental Information).

\section{Mode-dependent Configurational Free Energy Landscape from NMR conformers}
The statistical averaged structural parameters that enter the LE4PD are calculated from the configurational ensembles measured experimentally by NMR. Those conformers should represent the most probable configurational states of the protein. If they are complete, they should predict the dynamics of the protein through the use of the LE4PD equation. In our model all conformers are assumed to contribute equally to the full ensemble. Fluctuations around the local conformational states are imposed by applying a simple Gaussian Network Model.\cite{copperman2015predicting,bahar1997direct}  Once combined with the long-time predictions of the LE4PD the theory provides a realistic and computationally inexpensive method to predict the dynamics of proteins on a wide range of time scales, from the local fluctuations to the large, concerted, conformational transitions.\cite{copperman2015predicting} 

Because the determination of NMR conformers are affected by errors, we performed as a further test MD simulations of the same systems. Simulations were performed in explicit solvent: additional description of the simulation is contained in the Supplemental Information. For the PR95 protease monomer, simulations were performed starting from each of the twenty conformers in the NMR structure, resulting in a set of twenty production ensembles used as input to the LE4PD. Each simulation had $50 \ ns$ of production. For the IGF2R protein, the first conformer was chosen as the starting structure, and only one simulation with $150ns$ was performed.  For each trajectory the root mean square deviation (RMSD) was calculated and statistics were only collected in the equilibrated sections of the trajectory. The trajectories were also required to contain only reversible transitions, as monitored by the RMSD.

While explicit solvent atomistic classical molecular dynamics simulations are well-developed and can be quite accurate, building a dynamical model from NMR conformer structures using this method requires a few seconds on a standard desktop computer, while achieving molecular dynamic simulations with converged dynamics on the same timescale requires on the order of $10,000 $ - $100,000$ hours of processor time depending on the size of the protein.\cite{karplus2002molecular,adcock2006molecular} 

\subsection{Free energy Landscape and barrier crossing in the diffusive mode description}
The first three global modes of the $LE4PD$ in bond coordinates describe the rotations of the folded structure as the rotational diffusion tensor. For proteins which have an arbitrary folded structure, the full rotational diffusion equation of an anisotropic 3-dimensional body must be solved. This alters the relaxation of the three global modes describing the rotational relaxation in the inertial lab frame.\cite{favro1960theory,copperman2014coarse-grained} 

Internal modes from MD simulations present a complex free energy landscape with energetic barriers that increase in height with decreasing mode number: cooperative dynamics are described by the low index diffusive modes and requires the concerted transition over multiple local energy barriers. High index modes describe local motion and transitions.

\begin{figure}[htbp] 
\includegraphics[width=.85\columnwidth]{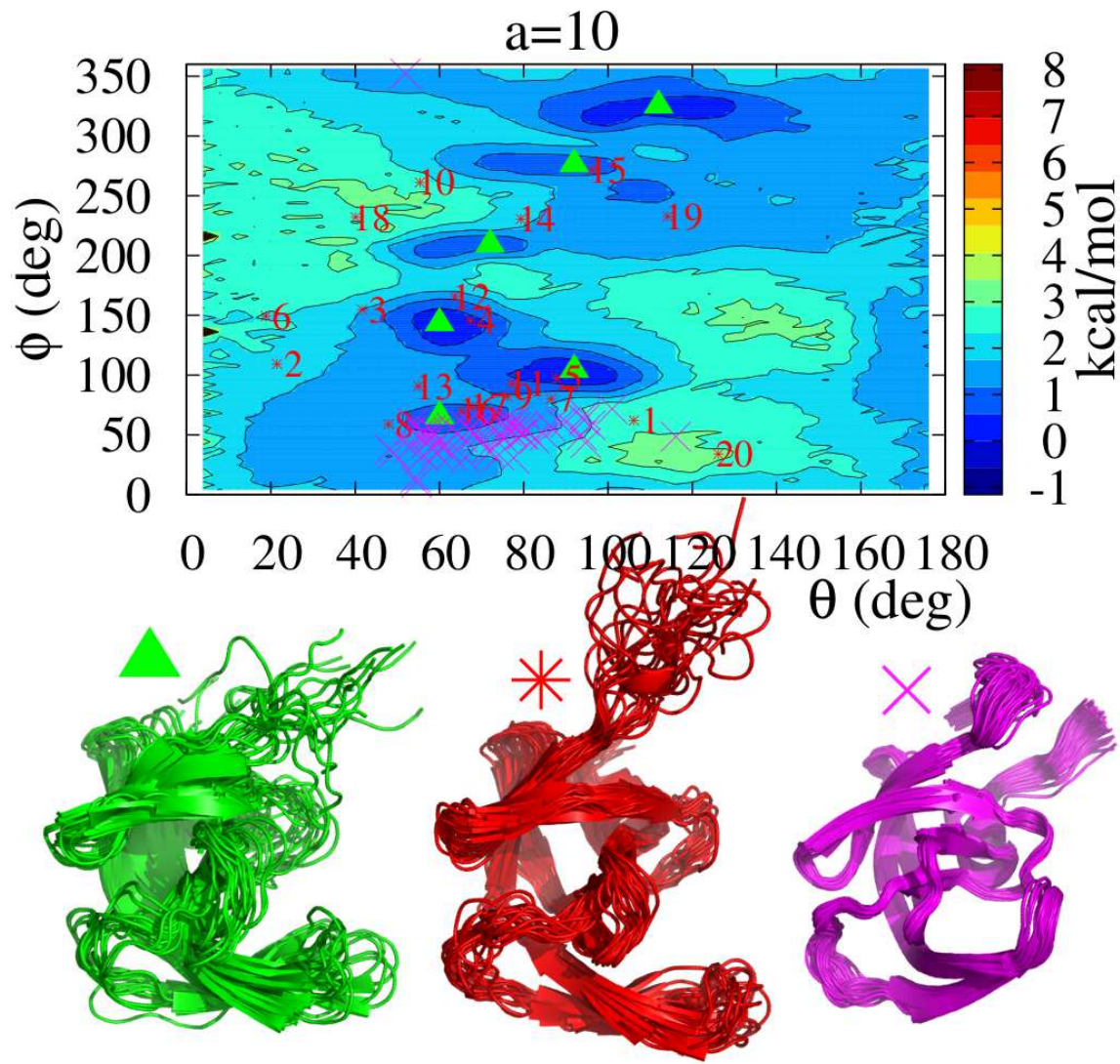}
\caption{Internal mode free energy surface (top) of the HIV protease monomer from MD simulation. Structural ensemble (below) of the monomer NMR conformer structures (red) and projected on the energy landscape as red stars, a dimerized and inhibited HIV protease (magenta), projected as magenta crosses, and average structures in minima from the MD simulations (green) and marked with green triangles.}
\label{modes}
\end{figure}

As an example, Figure \ref{modes} displays the mode-dependent free-energy landscape calculated from molecular dynamics simulations of the free monomer construct PR95 of the HIV protease. Each diffusive mode obtained from the diagonalization of Eq. \ref{LE} is a vector defined by the linear combination of the bond vectors weighted by the eigenvectors of the product of matrices $\textbf{LU}$, as $\vec{\xi}_a(t)=\sum_{i}Q^{-1}_{ai}\vec{l}_i(t)$. In polar coordinates the vector is represented as $\vec{\xi}_a(t)= \{| \vec{\xi}_a(t)|, \theta_a(t), \phi_a(t) \}$. The most relevant changes in the diffusive mode free energy occur as the angles, expressed in the spherical coordinates, span the configurational space while the effective bond length, i.e. the distance between two $\alpha$-carbons, is very close to constant. For any diffusive mode $a$, the free energy surface is defined as a function of the spherical coordinate angles $\theta_a$ and $\phi_a$ as  $F(\theta_a,\phi_a)= -k_B T \log \left \{P(\theta_a,\phi_a) \right\}$,with $P(\theta_a,\phi_a)$ the probability of finding the diffusive mode vector having the given value of the solid angle. The free energy landscape presents for each mode a number of minima separated by energy barriers.

\begin{figure}[htbp] 
\includegraphics[width=.85\columnwidth]{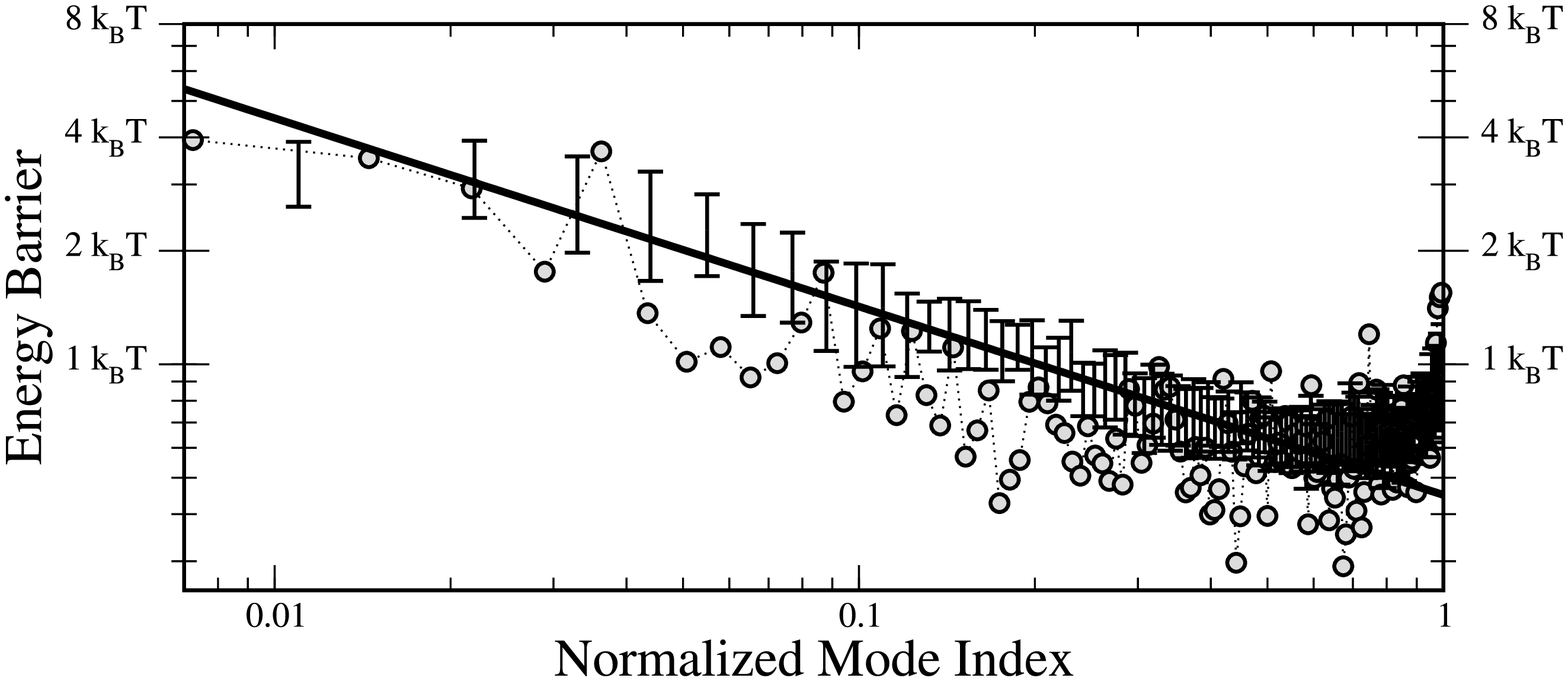}
\caption{The energy barriers in the modes calculated directly from the simulations of the HIV protease monomer PR95 at 293K plotted as the average over the 20 simulations with black error bars at 1 standard deviation, and from the simulation of the IGF2R protein at 273K (circles). The x-axis is the internal mode number normalized by the number of internal modes $N_p=N-4$ with $N$ the number of amino acids. The scaling form $E^\dag_a \propto (a-3)^{-.5}$ used in the simulation-free LE4PD model is plotted as a black line).}
\label{barriers}
\end{figure}

Figure \ref{barriers} shows the height of the median barrier measured in the mode-dependent free-energy energy landscape as calculated from the projection of molecular dynamics trajectory onto the LE4PD modes. The figure reports mode-dependent energy barriers for the HIV protease PR95 monomer and the IGF2R protein analyzed in this study. The behavior appears to follow a general scaling law. The height of the barrier scales inversely with the mode number as $E^\dag_a \propto (a-3)^{-.5}$ with $a$ the internal mode, so that small-index modes require highly cooperative transitions for the dynamics to take place. The local barrier measured in the configurational landscape of the highest diffusive mode is of the order of $k_BT$ and it is system dependent.

To account for the affect of the local energy barriers on the internal dynamics, the friction becomes mode dependent by assuming thermal activation over the mode-dependent energy barrier $\langle E^{\dag}_a \rangle$
\begin{eqnarray}
\overline{\zeta} \rightarrow \overline{\zeta}_a  \ exp[\langle E_a^{\dag} \rangle/(k_B T)] \ ,
\end{eqnarray}
leading to the slowing of the mode timescale $\tau_a = \frac{l^2 \overline{\zeta}}{3 k_B T \lambda_a }$ by
\begin{eqnarray}
\tau_a \rightarrow \tau_a  \ exp[\langle E_a^{\dag} \rangle/(k_B T)] \ .
\end{eqnarray}

This dynamical renormalization accounts for the scaling of the local barrier crossing, which is found to be general for all the proteins we studied.\cite{copperman2015predicting}  Once it is included in the calculation of the local dynamics the predictions of bond relaxation by LE4PD are found to be in quantitative agreement with experimental data of NMR $T_1$, $T_2$, and NOE relaxation across a set of 7 different proteins and 1864 site-specific measurements, with correlation coefficient $\rho>.9$, validating the LE4PD model.\cite{copperman2015predicting} As an example we report in the Supplemental Information the comparison between the predicted sequence-dependent NOE relaxation and experimental data of the IGF2R protein.

\section{Fluctuation Driven Dynamics of Binding}
\label{bio}
The LE4PD diffusive mode description provides useful information about enhanced fluctuations and the crossing of energy barriers in localized regions of the protein, as well as the lengthscale and timescale of these fluctuations. Through the analysis of the dynamics in diffusive modes, this study visualizes possible reaction pathways in mode coordinates for the pre-binding dynamics of the protein, and the time and space modulation of this dynamics after protein binding. Depending on the mode index, the dynamics of the unbound state involves the localized cooperative motion of groups of amino acids in specific parts of the protein, so that different spatial regions are dynamically active on different timescales. After binding, different active regions emerge with large amplitude uncorrelated motion for entropic compensation and more-localized enhanced fluctations indicating subsequent steps in the reaction mechanism. We report here the study of two proteins as an example. These proteins were selected because they have been studied structurally by NMR spectroscopy, and they are very diverse in that they perform different functions in different settings.

\subsection{HIV protease}
\begin{figure}[htbp] 
\includegraphics[width=.85\columnwidth]{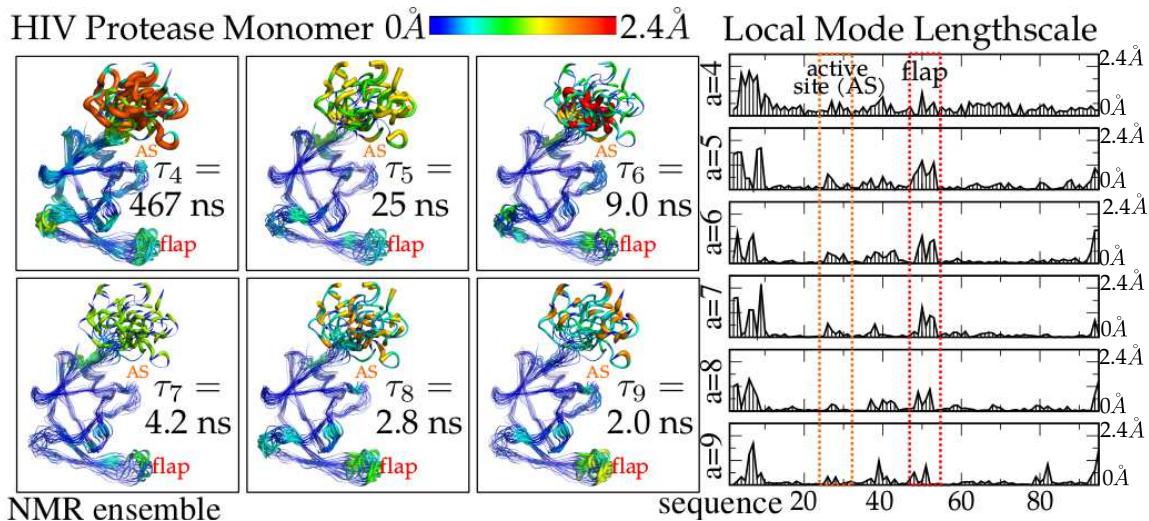}
\caption{Fluctuations of the free HIV protease monomer. Local mode lengthscale $L_{ia}$ along the primary sequence (Right panel), and projected as a rainbow gradient onto the ensemble of conformer structures 1Q9P (Left panel).}
% Flap motions, highlighted in red, are correlated with structural rearrangements of the active site highlighted in orange, amino acids $21$ to $33$.}
\label{proteasemon}
\end{figure}

As the first example, we consider  the dynamical model of the HIV protease monomer protein built from the NMR conformer ensemble 1Q9P.\cite{ishima2003solution} Figure \ref{proteasemon} shows the mode dependent localized fluctuations as a function of the amino acid position inside the protein primary sequence. In its active dimeric form aspartyl protease catalyzes the cleaving of the peptide chain for the separation of the protein products of the HIV genome.\cite{oroszlan1990retroviral} The PR95 monomer construct studied here is modified to stabilize the protease from self-cleavage, with five single-residue substitutions along the sequence and the deletion of the terminal residues $95$ - $99$. These C-terminal residues stabilize the dimer by forming antiparallel $\beta$-strands with the N-terminal residues $1$ - $5$.\cite{wlodawer1993structure} Without this stabilizing element, both the NMR relaxation rates and the ensemble of NMR conformers\cite{ishima2001folded,ishima2003solution} are consistent with an N-terminal tail that is highly flexible. Important conserved regions are the active site residues $22$ - $34$, the flap loop residues $47$ - $52$ which control access to the active site, and a region containing an $\alpha$-helix and part of the hydrophobic core residues $74$ - $87$.\cite{doron2005selecton} 

The dynamic localization is represented in Figure \ref{proteasemon} by projecting the mode amplitudes along the protein sequence, and making the radius and color of the tube rendering directly proportional to the extent of the local mode fluctuation (Left panel), with the related local mode fluctuation length defined as $L_{ia}^2=Q_{ia}^2\xi_a^2$ (Right panel). The analysis of the position inside the primary sequence of the large-amplitude internal modes shows that the slow modes are mostly localized along the flap loop and the terminal region. This is consistent with the role of the flap loop in HIV protease, mediating access to the active site. The LE4PD model suggests these structural fluctuations of the flap loop take place in the $1\ ns$ to $500 \ ns$ regime, in the unbound protein monomer. These motions are strongly correlated with smaller-scale conformational transitions in the active site, i.e. the sequence between amino acids $21$ and $33$, and the region around amino acid 40. The active site region, the flap region, and the $\alpha$-helix and hydrophobic core region $74$ - $87$ are conserved regions in the family of HIV protease proteins.

For the dimerized and inhibited HIV protease\cite{yamazaki1996three}, the structures from both of the proteins that make up the dimer are selected to be in the set of starting configurations for the LE4PD evaluation of the protein dynamics. The 28 solution structures in the ensemble lead to 56 total protein configurations for this analysis. The diffusive mode representation of this model shows that the enhanced mobility of the flap loop and the terminal regions observed in the HIV monomer are not present in the dynamics of the dimerized and inhibitor-bound molecular complex. 
  
\begin{figure}[htbp] 
\includegraphics[width=.85\columnwidth]{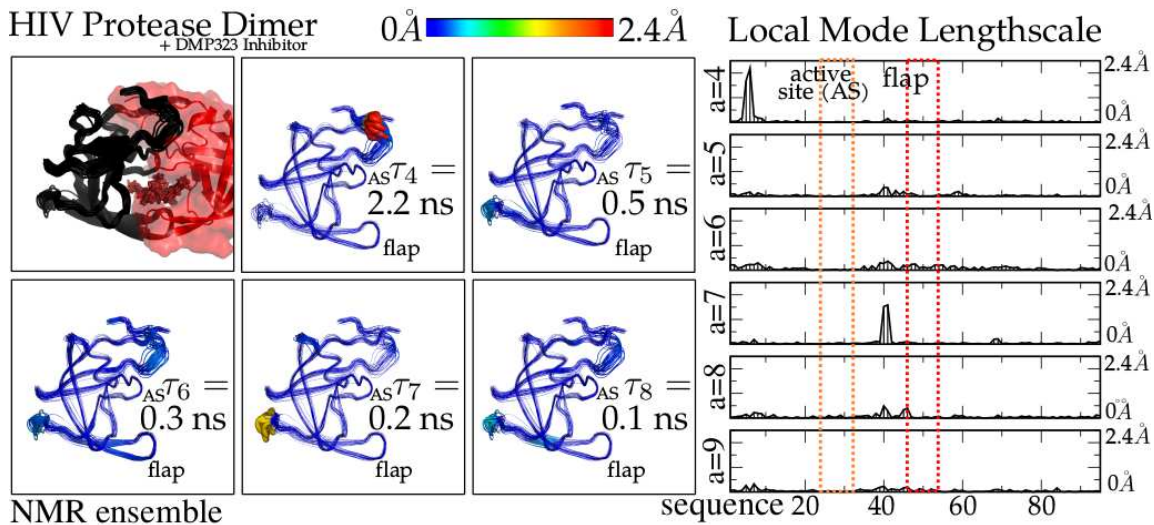}
\caption{Fluctuations of the dimerized and inhibitor-bound HIV protease. Local mode lengthscale $L_{ia}$ along the primary sequence (Right panel), and projected as a rainbow gradient onto the ensemble of conformer structures 1BVE (Left panel).}
% The protein in its bound form is highly structured and less flexible than in the monomer form. However, new flexibility arises around the P39:G40 amino acids, which may represent a subsequent step in the reaction mechanism and help compensate for the loss of configurational entropy of the flap and terminal regions due to binding.}
\label{proteasedim}
\end{figure}

As shown in Figure \ref{proteasedim} the large-amplitude slow modes observed in the monomer, are now trapped in the dimerized and inibitor-bound protein. The C-terminal and N-terminal tails which were previously disordered are now locked into a $\beta$-sheet structure, while motions of the flap loop and active site are inhibited by the binding of the DMP323 inhibitor. This indicates that the binding of the second protein and of the inhibitor efficiently trap specific structural minima in the ensemble of observed monomer conformers. This can be directly observed by projecting the dimerized and inhibitor bound protein structures onto the mode free energy landscape calculated from molecular dynamics simulations of the free monomer. It is clear in Figure \ref{modes} that the bound form is localized to specific regions and minima on the monomer energy landscape. Figure \ref{modes} also shows how the NMR configurational ensemble is sampling configurational states that are only partially consistent with the most stable states sampled in the MD simulation.

Interestingly, in Figure \ref{proteasedim} new fluctuations emerge that are localized to almost single-residue regions in previously more stable areas, such as the large-amplitude fluctuations of P39:G40 in the $200$ picosecond regime. In general large-amplitude short-time fluctuations help compensate with an increase of entropy for the loss of configurational entropy of the highly ordered complex. However, the well defined new peak that emerges here in the seventh mode, characterizing localized dynamics in the region of amino acid $40$, suggests a following step in the reaction pathway that would involve these amino acids. This is consistent with NMR relaxation experiments of the HIV protease/DMP323 inhibitor complex which found unusually large and fast motion at residue 40 as well, which was speculated to be involved involved in the release of the reaction product after protease activity.\cite{nicholson1995flexibility} Mutagenesis studies of the HIV protease found that non-conservative mutation of residue 40 resulted in loss of protease activity, even though it was separated from other highly conserved regions.\cite{loeb1989complete}  While this loop shows some flexibility in the free monomer,\cite{ishima2001folded} the LE4PD model shows an interesting form of dynamical allostery where the timescale of the local fluctuation becomes faster upon binding.

\subsection{IGF2R protein domain 11}
\begin{figure}[htbp] 
\includegraphics[width=.85\columnwidth]{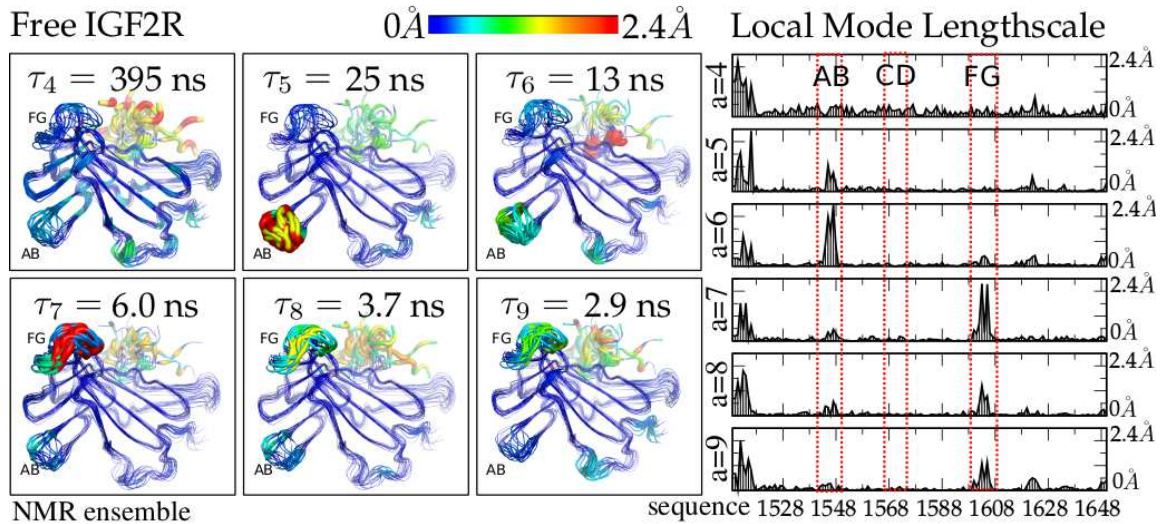}
\caption{Fluctuations of the free IGF2R domain 11 protein. Local mode lengthscale $L_{ia}$ along the primary sequence (Right panel), and projected as a rainbow gradient onto the ensemble of conformer structures 2M6T (Left panel).}
% Large-amplitude fluctuations of the binding site proceed from the AB loop at long timescale, to the FG loop at short timescale.}
\label{IGF2R}
\end{figure}

As a second example, we study the IGF2R protein domain 11 which is responsible for binding and regulating levels of IGF2 at the cell surface. Williams et al.\cite{williams2007structural} obtained relaxation and solution structures of this domain, and developed a binding model with the IGF2 protein where they suggested a dynamical role for two primary loops flanking a hydrophobic binding pocket. When analyzing the localization of the LE4PD modes on the protein sequence, we see that while the first internal mode primarily involves independent fluctuations of the unstructured C-tail, the second and third internal mode are localized highly on the AB and FG loops implicated in the binding model proposed by Williams et al. As can be seen in Figure \ref{IGF2R} these modes suggest largely independent, uncorrelated motion of these loops in the tens of nanosecond regime. The dynamics move from the AB loop to the FG loop on the $3 \ ns \ - \ 30 \ ns$ timescale, suggesting a sequence of steps in the binding pathway of the protein. Analysis of IGF2R sequences across species suggested that the gain of IGF2 binding affinity was accompanied by several distinct gain of function mutations.\cite{killian2000m6p} Most of these mutations were located in the IGF2R:IGF2 binding region, particularly residues 1544-1545 on the AB binding loop and residue 1600 on the FG loop.

\begin{figure}[htbp] 
\includegraphics[width=.85\columnwidth]{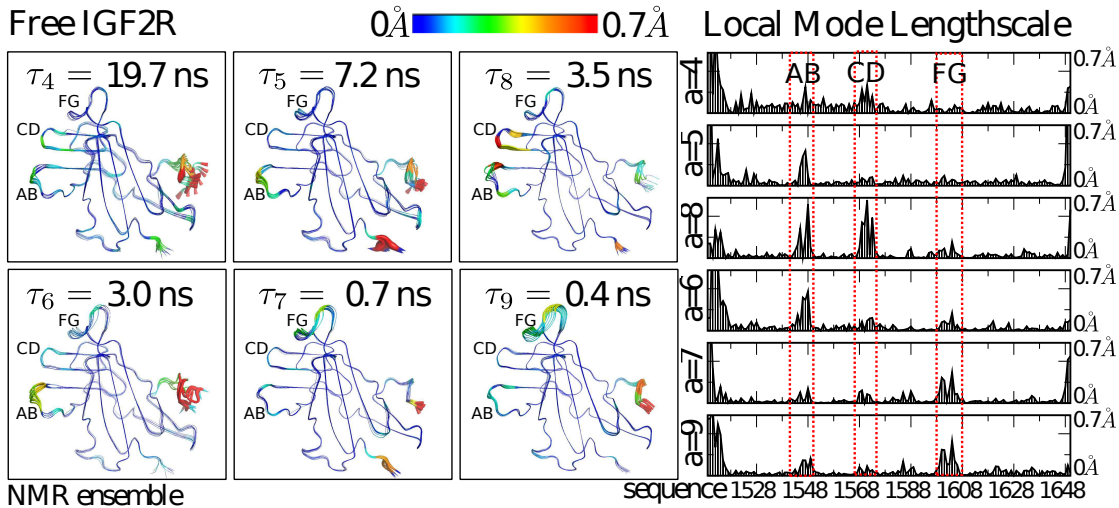}
\caption{Fluctuations of the free IGF2R domain 11 protein. Local mode lengthscale $L_{ia}$ along the primary sequence (Right panel) from the LE4PD model built from the MD simulation ensemble, and projected as a rainbow gradient onto the ensemble of structural minima in the mode (Left panel).}
% Large-amplitude fluctuations of the binding site proceed from the AB loop, to the CD loop, to the FG loop.}
\label{IGF2RMD}
\end{figure}

The dynamics presented so far for the IGF2R protein domain in the unbound and bound configurations is calculated using the LE4PD theory with input from the experimental structures measured in NMR. However, further calculations were performed with the LE4PD theory starting from molecular dynamics simulations. The simulation collected a $100 \ ns$ long trajectory with stable RMSD. Results for the mode dependent dynamics are reported in Figure \ref{IGF2RMD}.  The generated MD ensemble does not span the full range of dynamics observed in the NMR structural ensemble, which is not surprising since the variability in the NMR structural ensemble suggested mode dynamics in the $\sim \ 400 \ ns$ timescale, while the simulations are more limited in the timescale that they cover. 

Interestingly the simulation shows also a smaller lengthscale process propagating from the AB loop to the FG loop in the $1 \ ns \ - \ 20 \ ns$ timescale as can be seen in Figure \ref{IGF2RMD} in agreement with predictions from the NMR conformer ensemble. Furthermore, a new dynamical mode emerges in this intermediate timescale, which consists of correlated motion between the AB and the CD loops. This motion is not present in the NMR structural ensemble. However, considering the different timescales of the two calculations, the information collected from the two ensembles is consistent and complementary. 
 
\begin{figure}[htbp] 
\includegraphics[width=.85\columnwidth]{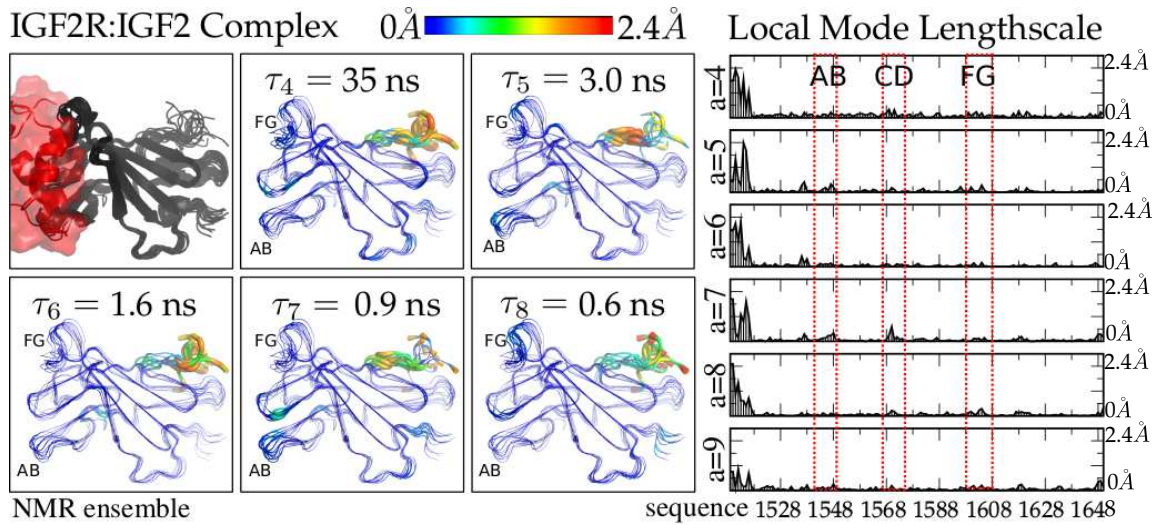}
\caption{Fluctuations of the bound IGF2R domain 11 protein. Local mode lengthscale $L_{ia}$ along the primary sequence (Right panel), and projected as a rainbow gradient onto the ensemble of conformer structures 2L29 (Left panel).}
% The tails of the protein are still mobile, but these fluctuations do not cooperatively propagate to the binding loop region of the protein.}
\label{IGF2:IGF2R}
\end{figure}

Upon binding, the dynamics of the loops of IGF2R domain becomes quite different. Starting from the IGF2R:IGF2 bound form from PDB 2L29\cite{williams2012exon} the dynamics of the three binding loops become quenched to a large extent. As can be seen in Figure \ref{IGF2:IGF2R}, in the bound state these loops maintain a small amount of flexibility but lack the cooperative long-time processes observed in the unbound protein. The tails of the protein are still mobile, but these fluctuations do not cooperatively propagate to the binding loop region of the protein. 

\section{Conclusions}

Residue specific localized dynamics involves the transition between, and the sampling of, multiple conformations. Those are localized in mobile parts of the protein that are mostly loops and terminal regions.  Local flexibility of the protein is needed for a conformational selection model of ligand binding.\cite{monod1965nature} Rather than these metastable states reflecting the endpoints of ligand bound conformations, this conformational diversity ensures the presence of energetically efficient binding pathways, and  the dynamics of motion between these metastable states.

Local dynamics are investigated with the LE4PD approach, a coarse-grained description that accounts for local semi-flexibility, energy barriers, hydrodynamics, and the chemical structure of the protein.\cite{copperman2015predicting} In its diffusive mode picture, the LE4PD defines  cooperative motion in localized regions of the protein. The relationship between cooperativity and energy barriers implies a sequence of relevant steps occurring to support the biological processes. In this way, the LE4PD theory has the ability to predict the emergence of localized fluctuations where mobility is enhanced at a given length and timescale, starting from the protein structural ensemble.

The most relevant information for the kinetics of binding is provided by the low index modes of motion as these large-amplitude, slow modes of motion identify cooperative fluctuations which are involved in the dynamics of recognition and binding of proteins to substrates. These slow dynamics are deemed relevant to the kinetic selection of protein binding partners as it detects regions where amino acids move cooperatively. The dynamics for these modes are characterized by relatively small barriers that are easily crossed in the nanosecond to several microsecond timescale that is relevant to protein recognition dynamics. As a consequence, the barriers can act as a switch that allows or forbids the transition between states through the precise modulation of the delicate balance of entropy and enthalpy. 

Regions of the activated dynamics are in general directly related to sequences that are conserved inside the family of proteins that perform a specific biological function. In previous work we showed how the predicted barrier-activated regions along the protein sequence corresponded to locations of signal transduction mechanisms of the CheY protein\cite{caballero2007theory} and ubiqitination linkage sites.\cite{copperman2014coarse-grained} Here, we show that an even more precise temporal and spatial description of the local dynamic activation can be obtained from the analysis of barrier crossing and mode dependent activated dynamics. A realistic and nearly quantitative picture of the processes involved in binding is obtained with minimal computational work from the LE4PD model constructed from ensembles of NMR conformers.

The solvent is a relevant player in these mechanisms of macromolecular binding, as they modulate the kinetics of the process by tuning the ordering of the solvent molecules, i.e. entropic contributions, and by breaking or forming hydrogen bonds, i.e. enthalpic contributions.\cite{levy2006water} In the LE4PD the effect of the solvent is implicitly included through friction, hydrodynamic interaction, and effective energy barriers. Because the input configurations are either calculated in an atomistic simulation with explicit representation of the solvent or from the experimental NMR conformers, a realistic description of the effect of the solvent is included in the projected dynamics of the LE4PD.

In general we observe that when the protein binds to a substrate, the original regions of energy activated motion are involved in the binding reaction mechanism and their dynamics becomes quenched. In the protein-substrate complex we observe the quencing of the motion in the single protein enhanced dynamics regions, and the emergence of new regions of higher flexibility in parts of the protein that are remote with respect to the binding interface. This emergence of new entropic states balances the reduction of entropy due to binding.  It is from the sophisticated balance of all these energetics, which to a large extent tend to compensate each other, that the physiological reaction pathways emerge in the biological mechanisms that regulate the function of proteins. These pathways must be evolutionarily tuned to avoid kinetic traps and ensure that binding partners can find their bound conformation. 
 
This study of the dynamics provides insight into protein recognition, which involves cooperative motion localized to active regions of the protein with fluctuations occurring over a hierarchy of length and timescales. Slow, correlated, spatially localized, fluctuations display a dynamical pathway which is relevant to the biological mechanism and function, while fast, uncorrelated fluctuations indicate simple entropic compensation after protein binding.

\begin{acknowledgments}
This work was supported by
the National Science Foundation Grant CHE-1362500. This work used the Extreme Science and Engineering Discovery Environment (XSEDE), which is supported by National Science Foundation grant number ACI-1053575.
\end{acknowledgments}

%\bibliography{ref_jcp}

\renewcommand{\thefigure}{S\arabic{figure}}
\section{Supplemental Material}
\subsection{Theory}
The Langevin equation formalism is derived  starting from the Liouville equation for the conservation of probability density in the phase space of the full atomistic system of the protein and solvent, and using projection operators to obtain an equation of motion for the chosen sites.\cite{zwanzig2001nonequilibrium} Here the chosen coarse-grained sites are the $\alpha$-carbon of each amino acid in the protein primary sequence. To obtain a linear Langevin equation we take the coordinates tracing the backbone configuration of the protein to be complete of the relevant slow configurational degrees of freedom, and neglect system memory. Inertial terms may be discarded as a protein in aqueous solution is safely in the overdamped limit. The intramolecular distribution around the folded state is assumed to be Gaussian, and the parameters in the distribution are directly obtained from the starting configurational ensemble.\cite{caballero2007theory,guenza2008theoretical}
The coarse-grained LE4PD represents the balance of viscous dissipation with the entropic restoring force and a random Brownian force due to the random collisions of the coarse-grained protein with the fast-moving projected atoms belonging to solvent, ions, and the protein. The time evolution of the coordinate of the coarse-grained site $i$ is well-described by the following equation 
\begin{eqnarray}
\overline{\zeta}\frac{\partial\vec{R}_i(t)}{\partial t}= - \frac{3 k_B T}{l^2} \sum_{j,k} H_{ij}A_{jk}\vec{R}_k(t)+\vec{F}_i(t) \ ,
\label{LE1}
\end{eqnarray}
 where $k_B$ is the Boltzmann constant, $T$ is the temperature, $l^2$ is the squared bond distance, and $\overline{\zeta}$ is the average monomer friction coefficient, defined as $\overline{\zeta}= N^{-1} \sum_{i=1}^{N} \zeta_i$, with $\zeta_i$ the friction of the monomer $i$. $\vec{F}_i(t)$ is a delta-correlated random force due to projecting the system dynamics onto the coarse-grained sites, where fluctuation-dissipation requires $\langle F_{i\alpha}(t) F_{j\beta}(t')\rangle = 2 k_B T \zeta_i \delta(t-t') \delta_{i,j}\delta_{\alpha,\beta}$ where $\alpha,\beta$ are cartesian indices.
 Eq. \ref{LE1} is the well-known Rouse-Zimm equation for the dynamics of polymers in solution.\cite{bixon1978optimized,doi1986theory}

%We assume a well-folded state where site-site correlations are Gaussian in nature, and the potential of mean force is derived from the standard structural $\mathbf{A}$ matrix built from the set of pairwise bond correlations.
To obtain an effective linear description we assume a well-folded state where site-site correlations are Gaussian in nature. The structural force matrix $\mathbf{A}$ defines the effective mean-force potential, $V(\{\vec{R}\})=\frac{3 k_B T}{2 l^2}\sum_{i,j=1}^{N}A_{ij}\vec{R}_i\cdot\vec{R}_j$, which has been successfully adopted in theories of protein folding to describe the final state of the folding process.\cite{portman1998variational} The $\mathbf{A}$ matrix is calculated as 
\begin{eqnarray}
\label{matrixA}
\mathbf{A}= \mathbf{M}^T \left( \begin{array}{cc}
0 &\mathbf{0} \\
\mathbf{0} & \mathbf{U} \\ \end{array} \right) \mathbf{M} \ ,
\end{eqnarray}
 where $\mathbf{M}$ is the matrix that defines the center of gyration and the connectivity between sites, $\sum_{j}M_{ij}\vec{R}_j=\vec{l}_i$. In a protein the $\alpha$-carbons are connected linearly, so that for $i > 1$ the matrix is defined as $M_{i,i-1}=-1$ and $M_{i,i}=1$, with $i=2, . . . , N$, while $M_{1,i}=1/N$ for the first row, and $M_{i,j}=0$ otherwise. The $\mathbf{U}$ matrix is the bond correlation matrix with $(\mathbf{U}^{-1})_{ij}=\frac{\langle \vec{l}_i \cdot \vec{l}_j \rangle}{\langle |\vec{l}_i | \rangle \langle | \vec{l}_j | \rangle }$.

The matrix $\textbf{H}$ is the 
%preaveraged
hydrodynamic interaction matrix, which describes the interaction between protein sites occurring through the liquid, described as a continuum medium. While it is standard to utilize hydrodynamical models to obtain the translational and rotational dynamics of proteins,\cite{garcia2000calculation} the contribution of hydrodynamical effects to protein internal motion is generally neglected. While this may be justified for very localized motion, in general the non-local hydrodynamic coupling alters the timescale and nature of the large-amplitude highly correlated internal motion and cannot be neglected.\cite{granek2011proteins,copperman2014coarse-grained} To maintain an effective linear description, the hydrodynamic interaction must be preaveraged. While the derivation of the hydrodynamic interaction utilizes the Oseen tensor following the general Rouse-Zimm treatment of polymer chains in dilute solution,\cite{doi1986theory} other methods such as the Rotne-Prager interaction tensor reduce to the same form upon preaveraging over the isotropic equilibrium distribution.\cite{rotne1969variational} The elements in the matrix of the hydrodynamic interaction are defined as 
\begin{eqnarray}
H_{ij}=\frac{\overline{\zeta}}{\zeta_i}\delta_{ij}+(1-\delta_{ij})\overline{r}^w \langle\frac{1}{r_{ij}}  \rangle  \ . 
\end{eqnarray}
where $\overline{r}^w= N^{-1} \sum_{i=1}^{N} r^w_i$ is the average hydrodynamic radius which is defined below. This is a perturbative hydrodynamic interaction accounting for the nature of the amino acid primary structure as a heteropolymer made up of building blocks of different chemical types, propagating through the aqueous solvent but screened in the dense hydrophobic core. The site-specific friction parameters, $\zeta_i$, are obtained by calculating the solvent-exposed surface area, and calculating the total friction of the $i_{th}$ site via a simple extension of Stoke's law as
\begin{eqnarray}
\zeta_i=6\pi(\eta_w r^w_i + \eta_p r^p_i) \ . 
\label{Stoke}
\end{eqnarray}
Here $\eta_w$ and $r^w$ denote, respectively, the viscosity of water and the radius of a spherical bead of identical surface area as the solvent-exposed surface area of the residue, the hydrodynamic radius\cite{caballero2007theory}, while $r^p$ denotes the hydrodynamic radius related to the surface not exposed to the solvent. The internal viscosity is $\eta_p$, which we approximated in our previous work to be related to the water viscosity rescaled by the local energy-barrier scale $\sim k_B T$.\cite{copperman2014coarse-grained,sagnella2000time} The largest possible value of $\overline{r}^w$ that maintains a positive definite solution of the matrix diagonalization is adopted to avoid the well-known issue with the preaveraging of the hydrodynamic interaction in dense systems.\cite{zwanzig1968validity} For example, in the application of the model to HIV protease, the calculated $\overline{r}^w=2.28\AA$ is very close to the adopted value of $\overline{r}^w=2.23\AA$, which avoids negative eigenvalues.

Because we focus only on the bond orientational dynamics and not translation, in the interest of a simpler notation we separate out the zeroth order translational mode from the internal dynamics. Following the same notation introduced for the orientational dynamics of star polymers,\cite{guenza1992reduced} 
we define $\mathbf{a}$ as the $\mathbf{M}$ matrix after suppressing the first row used to define the center of mass, and define $\mathbf{L}=\mathbf{a} \mathbf{H} \mathbf{a}^T$. The orientational  Langevin equation governing the bond dynamics is 
\begin{eqnarray}
\frac{\partial\vec{l}_i (t)}{\partial t}= - \sigma  \sum_{j,k} L_{ij}U_{jk}\vec{l}_k(t)+\vec{v}_i(t) \ ,
\label{LE}
\end{eqnarray}
 with $i,j=1, ..., N-1$, and where $\sigma=3 k_B T/(l^2\overline{\zeta})$, and $\vec{v}_i(t)$ is the random delta-correlated bond velocity. 
%Eq. \ref{LE} represents a set of $N-1$ first-order coupled differential equations, which are solved by diagonalization of the matrix product $\mathbf{LU}$ with $\mathbf{Q}$ the eigenvectors and $\mathbf{\lambda}$ the eigenvalues. In the linearly independent normal modes $\vec{\xi_a}(t)=\sum_{i}Q^{-1}_{ai}\vec{l}_i(t)$. Like the set of bond vectors $\vec{l}_i(t)$ the set of normal coordinates $\vec{\xi}_a(t)$ defines the instantaneous conformation of the macromolecule. The normal mode basis spans the same space as the bond vector basis with the added benefit of linearity: $\langle\vec{\xi}_a (t) \cdot\vec{\xi}_b (t) \rangle=\delta_{ab}  l^2/\mu_a$ with  the eigenvalues of the bond correlation matrix given by $\sum_{i,j}Q^{-1}_{ai}U_{ij}Q_{jb}=\delta_{ab}\mu_a$.

\subsection{Molecular Dynamics Simulations}
Simulations were performed in explicit solvent using the spc/e water model. We utilized the AMBER99SB-ILDN\cite{lindorff2010improved} atomic force field for proteins and the GROMACS\cite{gromacs1,gromacs2,gromacs3,gromacs4} molecular dynamics engine on the TRESTLES supercomputer at San Diego.\cite{towns2014xsede} All system conditions, e.g. temperature and salt concentration, were set to reproduce the experimental conditions of the NMR solution structure. The systems were solvated and energy minimized, and then underwent a $500 \ ps$ tempering and equilibration routine including pressure coupling. The production simulations were performed in the canonical ensemble, using a velocity rescaling thermostat.\cite{bussi2007canonical}

\subsection{Testing the Accuracy of the Model}
\begin{figure}[htb] 
\includegraphics[width=.85\columnwidth]{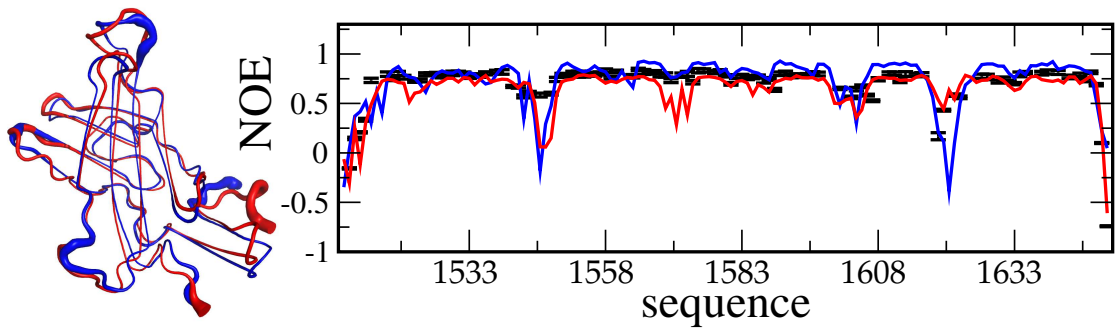}
\caption{Right: Site-specific measurements of heteronuclear Overhauser effect (NOE) along the backbone of the IGF2R protein. The experimental values (black) are compared with theoretical predictions calculated from the LE4PD dynamical model using as an input the MD generated structural ensemble (red), and from the LE4PD dynamical model using as an input the NMR solution structure ensemble (blue). Left: Average configuration from the MD simulation ensemble (red) and from the NMR structural ensemble (blue). The thickness of the ribbon is accurate to the local orientational fluctuations of the conformational distribution.}
\label{NOE}
\end{figure}

As an example we report the comparison between the predicted sequence-dependent NOE relaxation and experimental data of the IGF2R protein. We compare the LE4PD dynamical models created from NMR solution structure ensembles and the  LE4PD dynamical models created from MD generated structural ensemble, with the data of NMR backbone relaxation. Figure \ref{NOE} shows that both LE4PD models are quantitatively consistent with the experimental data. Similar results were found across a set of 7 different proteins and 1864 site-specific measurements, with correlation coefficient $\rho>.9$, validating the LE4PD model.\cite{copperman2015predicting}

%\bibliography{ref_jcp}

\end{document}